\documentclass[floats,twocolumn]{revtex4}

\usepackage{amsmath}
\usepackage{epsfig}

\newcommand{\g}[1]{#1}             
\newcommand{\n}[1]{(#1)}           
\newcommand{\z}[1]{[#1]}           
\newcommand{\p}[1]{\{#1\}}         
\newcommand{\pz}[1]{[\{#1\}]}      

\renewcommand{\g}[1]{\mathbf{#1}}    
\renewcommand{\n}[1]{\underline{#1}} 
\renewcommand{\z}[1]{{#1}}    
\renewcommand{\p}[1]{\mathit{#1}}    
\renewcommand{\pz}[1]{\mathit{#1}}   

\begin{document}

\title{Systematic comparison of force fields for microscopic simulations of NaCl 
      in aqueous solutions: Diffusion, free energy of hydration
      and structural properties }

\author{Michael Patra}
\author{Mikko Karttunen}
\affiliation{Biophysics and Statistical Mechanics Group,
Laboratory for Computational Engineering, 
Helsinki University of Technology, P.\,O. Box 9203, FIN-02015 HUT, Finland}

\begin{abstract}
In this paper we compare 
different force fields that are widely used (Gromacs, Charmm-22/x-Plor,
Charmm-27, Amber-1999, OPLS-AA) 
in  biophysical simulations containing aqueous NaCl.  
We show that the uncertainties
of the microscopic parameters of, in particular, sodium and, to a lesser extent,
chloride translate into large differences in the computed radial-distribution
functions. This uncertainty reflects the incomplete experimental knowledge
of the structural properties of ionic aqueous solutions at finite molarity.
We discuss possible implications on the computation of
potential of mean force and effective potentials.
\end{abstract}

\maketitle

\section{Introduction}

The presence of water is characteristic to all biological systems. 
For any simulational study of biophysical phenomena, a proper description of water
is a necessity. It is widely known that modelling water is a difficult task, 
and due to this many different models have been developed.
In addition to water, the description of ions, in particular Na$^+$ and Cl$^-$, 
is essential for biophysical systems. A simple example is the physiological liquid
in the human body containing about $0.8~\mathrm{mol}$ salt, i.e., of the 
order of one ion pair per $100$ water molecules. 

Today, virtually all simulations use one of the water models either from 
the TIP or the SPC series, 
a recent review of water models is provided by Wallqvist and Mountain~\cite{wallqvist:99a}.
For water, the influence of aspects such as density, treatment of
long-range electrostatics and the choice of force-field
have been studied intensively, see e.g. Ref.~\onlinecite{mark:02a} for a recent study. 
However, no systematic studies, to the
authors' knowledge, exist for ionic force fields.
On one hand that is very surprising since various studies indicate that the
force field can have a significant effect on the system
properties even in the case of an implicit solvent~\cite{dang:92a}. 
In addition, it also known from experiments that the properties of ionic 
solutions may be significantly influenced even by small amounts 
of heavy isotopes of the ions~\cite{chakrabarti:95a}.

On the other hand, a practical obstacle for a systematic comparison has been the
evaluation of long-range electrostatics. In the past, there has been much
work to include the effects of Coulombic interactions~\cite{perera:94a,buono:94a,
zhu:92a,hummer:93a,hummer:94b,restrepo:94a,chialvo:95a,lyubartsev:95a}
without performing the computationally costly Ewald summations.
Present day computational resources allow one to treat electrostatics 
properly~\cite{sagui:99a} and in, e.g., the case of lipid bilayers,
this is a very important matter~\cite{patra:03a}. 

In this article we compare various commonly used force fields and a parametrisation
for sodium chloride in combination with different commonly used water models.
For NaCl we used Gromacs, Charmm-22/x-plor, Charmm-27, Amber-1999 and OPLS-AA force fields,
and the parametrisation by Smith and Dang~\cite{smith:94b}.
The water models used for the aqueous solution were SPC, SPC/E, TIP3P and TIP4P.
In previous studies some static properties such as radial distribution 
functions have been studied,
but in each case for one particular choice of force 
field 
only~\cite{zhu:92a,hummer:93a,hummer:94b,restrepo:94a,chialvo:95a,lyubartsev:95a,kovalenko:00a}. 
In addition, the effects of 
temperature~\cite{koneshan:00a} and salt concentration~\cite{chowdhuri:01a}
have recently been studied. To our knowledge, there is no systematic study regarding the
effects of the force field and this study aims to fill that gap.

When developing empirical force fields, one matches certain  experimental
quantities with their counterparts as determined by a numerical simulation.
The choice is determined by the availability of high quality experimental data
and the physical significance of that quantity. The parameters of a force field
thus depend critically on the choice of quantities that are being compared. 
Force fields are typically optimised for macroscopic parameters like the
Gibbs free energy of hydration (i.\,e., placing an ion into a shell of water
should give
the same lowering of energy in the simulation as in an experiment) that are
important thermodynamic quantities and at the same time can be measured to a
high accuracy. 

The thermodynamic properties are best complemented by structural properties.  A
natural way  to quantify them is by using distribution functions of which  the
radial-distribution function $g_{ij}(r)$ is the most commonly used,  where
$r=|\vec{r}_i-\vec{r}_j|$ stands for the separation between a particle of
component $i$ and of component $j$.  It gives the probability of finding two
particles at some distance $r$, taking account of density and geometric
effects, and can be formally related to the potential of mean force between two
particles~\cite{hansen:86}.  Aqueous NaCl is characterised by six different
radial-distribution functions (since there are three components) but three of
them cannot be determined experimentally at all, and of two only limited
experimental information is available. We will return to this later in this
paper.

All force fields discussed in this paper have been parameterised using the
available experimental information on aqueous NaCl and therefore reproduce those
experimental values reasonable well. In contrast, structural properties
without known experimental values vary significantly between force fields.
This uncertainty has become a significant problem recently since structural
properties are essential in force field
development for coarse-grained systems~\cite{meyer:00a,plathe:02a}.

\section{Force Fields}
\label{secForceFields}

In this article we compare different force fields. We restrict ourselves
to force fields in the traditional sense, i.\,e., they have to be available
in an electronic form and cover a wide range of systems. We
specify the precise files used to obtain the parameters for our
simulations. This information is relevant  since often 
these parameters vary slightly between different
sources even for the same force field.

It should be noted that all of the force fields tested here
were originally developed with the aim
of describing proteins and nucleic acids. Description of  ions is thus
only a small part of their capabilities. For comparison, 
we have included one hand-optimised set of parameters for NaCl only~\cite{smith:94b}.

The different force fields and files are the following:

\begin{description}

\item[Gromacs (``GROM''):] Force field included in Version 3.1.3 of
Gromacs~\cite{lindahl:01a}. Available at
\texttt{http://www.gromacs.org/\discretionary{}{}{}download/\discretionary{}{}{}index.php};  file
\texttt{ffgmxnb.itp}. The TIP4P water model for Gromacs is available at 
\texttt{http://www.gromacs.org/\discretionary{}{}{}pipermail/\discretionary{}{}{}gmx-users/\discretionary{}{}{}2001-November/\discretionary{}{}{}000152.html}.
For the systems discussed in this paper, i.\,e., water and NaCl, 
the Gromacs force field is identical to the Gromos-96 force field. Since
Gromacs is the fastest MD program available, its default force field
is used increasingly often.

\item[X-Plor/Charmm-22 (``XPLR''):] Force field from x-plor distribution
3.851, available at \texttt{http://atb.csb.yale.edu/\discretionary{}{}{}xplor/}; file
\texttt{parallh22x.pro}. While this force-field is labelled as Charmm-22, the original
Charmm-22 force field~\cite{mackerell:98a} does not include ions, but they are
included only in the x-plor distribution. X-plor~\cite{brunger:92} is one of the most 
versatile non-commercial programs for protein simulations but is only
able to use this force field.

\item[Charmm-27 (``CH27''):] Available at\\
\texttt{https://rxsecure.umaryland.edu/\discretionary{}{}{}research/\discretionary{}{}{}amackere/\discretionary{}{}{}research.html}; \\file
\texttt{par\_all27\_prot\_lipid.inp}. In comparison to Charmm-22, the more
recent  Charmm-27~\cite{foloppe:00a}  contains parameters for
ions in the files available at the website. Charmm-27 includes also other
improvements for the description nucleic acids.  For proteins, 
Charmm-22 is identical to Charmm-27.
The Charmm-27 ion parameters are credited
to Refs.~\onlinecite{beglov:94a} and~\onlinecite{roux:96a}.

\item[Amber-1999 (``AMBR''):] The complete force field distribution for the
Amber-1999 force field~\cite{wang:00a} is available at
\texttt{http://www.amber.ucsf.edu/\discretionary{}{}{}amber/\discretionary{}{}{}amber7.ffparms.tar.gz}.
We used parameter file \texttt{parm99.dat} and TIP4P water model
from \texttt{frcmod.tip4p}. New Amber-2002 force field includes
explicit polarisation terms, and thus falls outside the scope of this
comparison.

\item[OPLS-AA (``OPLS''):] The OPLS-AA force field~\cite{rizzo:99a} is the only
force-field in our list that is not part of a MD simulation
package. Hence, there  is no ``official'' file with the force field
parameters. We chose the one included with Gromacs
Version 3.1.4,  available at
\texttt{http://www.gromacs.org/\discretionary{}{}{}download/\discretionary{}{}{}index.php} in file
\texttt{ffoplsaanb.itp}. One should note that other sources exist,
e.\,g. 
\texttt{http://www.scripps.edu/\discretionary{}{}{}brooks/\discretionary{}{}{}charmm\_docs/\discretionary{}{}{}oplsaa-toppar.tar}.

\item[Smith-1994 (``SMIT''):] Hand-optimised set. Published in Ref.~\onlinecite{smith:94b}.

\end{description}

We performed simulations using the four standard water models, namely  the rigid
versions of SPC~\cite{berendsen:81a}, SPC/E~\cite{berendsen:87a},
TIP3P~\cite{jorgensen:83a,neria:96a} and TIP4P~\cite{jorgensen:83a}. For
computational efficiency, we did not use the flexible versions of the water
models. 
SPC/E and SPC differ only by the partial charges assigned to the atoms, so that the
Lennard-Jones parameters are identical and thus need to be specified only once. 

The computer-readable files from the force field distributions typically contain
one or more of the above water models. Whenever a water model was available in
this way, we took the parameters from that file. Otherwise, standard
parameters were used. This explains the (very)
small differences than can be seen in Tab.~\ref{tabParameter} between different
force field distributions for the same water model.

The assignment of partial charges for the ions is trivial, and for the
different water models it is well defined by the water model. The relevant
parameters, since they are different for each force-field, thus
are the ones describing Lennard-Jones interactions. They can be specified 
in different ways, the two most common ones being
\begin{equation}
        V(r) = \frac{c_{12}}{r^{12}} - \frac{c_6}{r^6} 
             = 4 \epsilon \left[ \left(\frac{\sigma}{r}\right)^{12} -
                \left(\frac{\sigma}{r}\right)^6 \right] \;,
        \label{eqconv}
\end{equation}
where the freedom of measuring energy in  kcal or 
kJ remains. In addition, another  common practise is 
not to specify all interaction parameters explicitly 
but to use the Lorentz-Berthelot  combination rules~\cite{leach:96}
\begin{equation}
        \epsilon = \sqrt{ \epsilon_1 \epsilon_2}
	 \;, \qquad
        \sigma = \frac{\sigma_1 + \sigma_2 }{2} \;,
        \label{eqKombination}
\end{equation}
where the indices 1 and 2 denote particles of type 1 and 2, respectively.
Table~\ref{tabParameter} lists the precise Lennard-Jones
parameters used in our simulations.
In addition, the table also indicates whether the parameter in question was
specified directly by the force field or had to be computed via
Eqs.~(\ref{eqconv}) and/or~(\ref{eqKombination}).

\begin{table*}
\small
\noindent%
\begin{tabular}{@{}lllll}
\hline
\multicolumn{5}{c}{{\normalsize\textbf{Gromacs}}} \\
\hline
Atom & $c_6~[\mathrm{kJ}\,\mathrm{nm}^6]$ 
        & $c_{12}~[\mathrm{kJ}\,\mathrm{nm}^{12}]$ & 
        $\epsilon~[\frac{\mathrm{kcal}}{\mathrm{mol}}]$ & $\sigma~[\mathrm{\AA}]$ \\
\hline
Cl &
      $\g{ 1.3804\cdot 10^{-2 }}$ & $\g{ 1.0691\cdot 10^{-4 }}$ &
      $\n{ 0.1064}$ & $\n{ 4.4480}$ \\
Na &
      $\g{ 7.2059\cdot 10^{-5 }}$ & $\g{ 2.1014\cdot 10^{-8 }}$ &
      $\n{ 0.0148}$ & $\n{ 2.5752}$ \\
O (S) &
      $\g{ 2.6171\cdot 10^{-3 }}$ & $\g{ 2.6331\cdot 10^{-6 }}$ &
      $\n{ 0.1553}$ & $\n{ 3.1655}$ \\
O (3) &
      $\g{ 2.4889\cdot 10^{-3 }}$ & $\g{ 2.4352\cdot 10^{-6 }}$ &
      $\n{ 0.1519}$ & $\n{ 3.1508}$ \\
O (4) &
      $\n{ 2.5543\cdot 10^{-3 }}$ & $\n{ 2.5145\cdot 10^{-6 }}$ &
      $\g{ 0.1549}$ & $\g{ 3.1540}$ \\
\hline
Cl---Na &
      $\g{ 9.9737\cdot 10^{-4 }}$ & $\g{ 1.4989\cdot 10^{-6 }}$ &
      $\n{ 0.0396}$ & $\n{ 3.3844}$ \\
Cl---O (S) &
      $\g{ 6.0106\cdot 10^{-3 }}$ & $\g{ 1.6778\cdot 10^{-5 }}$ &
      $\n{ 0.1286}$ & $\n{ 3.7524}$ \\
Na---O (S) &
      $\g{ 4.3426\cdot 10^{-4 }}$ & $\g{ 2.3523\cdot 10^{-7 }}$ &
      $\n{ 0.0479}$ & $\n{ 2.8551}$ \\
Cl---O (3) &
      $\g{ 5.8616\cdot 10^{-3 }}$ & $\g{ 1.6135\cdot 10^{-5 }}$ &
      $\n{ 0.1272}$ & $\n{ 3.7436}$ \\
Na---O (3) &
      $\g{ 4.2350\cdot 10^{-4 }}$ & $\g{ 2.2622\cdot 10^{-7 }}$ &
      $\n{ 0.0473}$ & $\n{ 2.8485}$ \\
Cl---O (4) &
      $\z{ 6.4856\cdot 10^{-3 }}$ & $\z{ 1.9559\cdot 10^{-5 }}$ &
      $\z{ 0.1284}$ & $\z{ 3.8010}$ \\
Na---O (4) &
      $\z{ 4.4243\cdot 10^{-4 }}$ & $\z{ 2.4446\cdot 10^{-7 }}$ &
      $\z{ 0.0478}$ & $\z{ 2.8646}$ \\
\hline
\end{tabular}
\hfill\begin{tabular}{lllll}
\hline
\multicolumn{5}{c}{{\normalsize\textbf{X-Plor / Charmm-22}}} \\
\hline
Atom & $c_6~[\mathrm{kJ}\,\mathrm{nm}^6]$ 
        & $c_{12}~[\mathrm{kJ}\,\mathrm{nm}^{12}]$ & 
        $\epsilon~[\frac{\mathrm{kcal}}{\mathrm{mol}}]$ & $\sigma~[\mathrm{\AA}]$ \\
\hline
Cl &
      $\n{ 1.5362\cdot 10^{-2 }}$ & $\n{ 9.3940\cdot 10^{-5 }}$ &
      $\g{ 0.1500}$ & $\g{ 4.2763}$ \\
Na &
      $\n{ 6.9284\cdot 10^{-4 }}$ & $\n{ 2.8663\cdot 10^{-7 }}$ &
      $\g{ 0.1000}$ & $\g{ 2.7297}$ \\
O (S) &
      $\p{ 2.6171\cdot 10^{-3 }}$ & $\p{ 2.6331\cdot 10^{-6 }}$ &
      $\p{ 0.1553}$ & $\p{ 3.1655}$ \\
O (3) &
      $\n{ 2.4913\cdot 10^{-3 }}$ & $\n{ 2.4366\cdot 10^{-6 }}$ &
      $\g{ 0.1521}$ & $\g{ 3.1506}$ \\
O (4) &
      $\p{ 2.5543\cdot 10^{-3 }}$ & $\p{ 2.5145\cdot 10^{-6 }}$ &
      $\p{ 0.1549}$ & $\p{ 3.1540}$ \\
\hline
Cl---Na &
      $\z{ 3.7899\cdot 10^{-3 }}$ & $\z{ 7.0028\cdot 10^{-6 }}$ &
      $\z{ 0.1225}$ & $\z{ 3.5030}$ \\
Cl---O (S) &
      $\pz{ 6.7840\cdot 10^{-3 }}$ & $\pz{ 1.8004\cdot 10^{-5 }}$ &
      $\pz{ 0.1526}$ & $\pz{ 3.7209}$ \\
Na---O (S) &
      $\pz{ 1.3689\cdot 10^{-3 }}$ & $\pz{ 8.9779\cdot 10^{-7 }}$ &
      $\pz{ 0.1246}$ & $\pz{ 2.9476}$ \\
Cl---O (3) &
      $\z{ 6.6331\cdot 10^{-3 }}$ & $\z{ 1.7393\cdot 10^{-5 }}$ &
      $\z{ 0.1510}$ & $\z{ 3.7134}$ \\
Na---O (3) &
      $\z{ 1.3342\cdot 10^{-3 }}$ & $\z{ 8.6187\cdot 10^{-7 }}$ &
      $\z{ 0.1233}$ & $\z{ 2.9402}$ \\
Cl---O (4) &
      $\pz{ 6.7132\cdot 10^{-3 }}$ & $\pz{ 1.7652\cdot 10^{-5 }}$ &
      $\pz{ 0.1524}$ & $\pz{ 3.7151}$ \\
Na---O (4) &
      $\pz{ 1.3513\cdot 10^{-3 }}$ & $\pz{ 8.7593\cdot 10^{-7 }}$ &
      $\pz{ 0.1245}$ & $\pz{ 2.9419}$ \\
\hline
\end{tabular}\\[3mm]
\noindent%
\begin{tabular}{lllll}
\hline
\multicolumn{5}{c}{{\normalsize\textbf{Charmm-27}}} \\
\hline
Atom & $c_6~[\mathrm{kJ}\,\mathrm{nm}^6]$ 
        & $c_{12}~[\mathrm{kJ}\,\mathrm{nm}^{12}]$ & 
        $\epsilon~[\frac{\mathrm{kcal}}{\mathrm{mol}}]$ & $\sigma~[\mathrm{\AA}]$ \\
\hline
Cl &
      $\n{ 1.0999\cdot 10^{-2 }}$ & $\n{ 4.8155\cdot 10^{-5 }}$ &
      $\n{ 0.1500}$ & $\n{ 4.0447}$ \\
Na &
      $\n{ 1.6169\cdot 10^{-4 }}$ & $\n{ 3.3284\cdot 10^{-8 }}$ &
      $\n{ 0.0469}$ & $\n{ 2.4299}$ \\
O (S) &
      $\p{ 2.6171\cdot 10^{-3 }}$ & $\p{ 2.6331\cdot 10^{-6 }}$ &
      $\p{ 0.1553}$ & $\p{ 3.1655}$ \\
O (3) &
      $\n{ 2.4912\cdot 10^{-3 }}$ & $\n{ 2.4364\cdot 10^{-6 }}$ &
      $\n{ 0.1521}$ & $\n{ 3.1506}$ \\
O (4) &
      $\p{ 2.5543\cdot 10^{-3 }}$ & $\p{ 2.5145\cdot 10^{-6 }}$ &
      $\p{ 0.1549}$ & $\p{ 3.1540}$ \\
\hline
Cl---Na &
      $\z{ 1.6169\cdot 10^{-3 }}$ & $\z{ 1.8611\cdot 10^{-6 }}$ &
      $\z{ 0.0839}$ & $\z{ 3.2373}$ \\
Cl---O (S) &
      $\pz{ 5.6117\cdot 10^{-3 }}$ & $\pz{ 1.2319\cdot 10^{-5 }}$ &
      $\pz{ 0.1526}$ & $\pz{ 3.6051}$ \\
Na---O (S) &
      $\pz{ 6.8542\cdot 10^{-4 }}$ & $\pz{ 3.2868\cdot 10^{-7 }}$ &
      $\pz{ 0.0853}$ & $\pz{ 2.7977}$ \\
Cl---O (3) &
      $\z{ 5.4847\cdot 10^{-3 }}$ & $\z{ 1.1892\cdot 10^{-5 }}$ &
      $\z{ 0.1510}$ & $\z{ 3.5976}$ \\
Na---O (3) &
      $\z{ 6.6750\cdot 10^{-4 }}$ & $\z{ 3.1500\cdot 10^{-7 }}$ &
      $\z{ 0.0845}$ & $\z{ 2.7903}$ \\
Cl---O (4) &
      $\pz{ 5.5514\cdot 10^{-3 }}$ & $\pz{ 1.2071\cdot 10^{-5 }}$ &
      $\pz{ 0.1524}$ & $\pz{ 3.5993}$ \\
Na---O (4) &
      $\pz{ 6.7619\cdot 10^{-4 }}$ & $\pz{ 3.2028\cdot 10^{-7 }}$ &
      $\pz{ 0.0852}$ & $\pz{ 2.7920}$ \\
\hline
\end{tabular}
\hfill\begin{tabular}{lllll}
\hline
\multicolumn{5}{c}{{\normalsize\textbf{Amber-1999}}} \\
\hline
Atom & $c_6~[\mathrm{kJ}\,\mathrm{nm}^6]$ 
        & $c_{12}~[\mathrm{kJ}\,\mathrm{nm}^{12}]$ & 
        $\epsilon~[\frac{\mathrm{kcal}}{\mathrm{mol}}]$ & $\sigma~[\mathrm{\AA}]$ \\
\hline
Cl &
      $\n{ 1.2170\cdot 10^{-2 }}$ & $\n{ 8.8431\cdot 10^{-5 }}$ &
      $\n{ 0.1000}$ & $\n{ 4.4010}$ \\
Na &
      $\n{ 6.3072\cdot 10^{-5 }}$ & $\n{ 8.5752\cdot 10^{-8 }}$ &
      $\n{ 0.0028}$ & $\n{ 3.3284}$ \\
O (S) &
      $\p{ 2.6171\cdot 10^{-3 }}$ & $\p{ 2.6331\cdot 10^{-6 }}$ &
      $\p{ 0.1553}$ & $\p{ 3.1655}$ \\
O (3) &
      $\n{ 2.4904\cdot 10^{-3 }}$ & $\n{ 2.4364\cdot 10^{-6 }}$ &
      $\n{ 0.1520}$ & $\n{ 3.1508}$ \\
O (4) &
      $\n{ 2.5534\cdot 10^{-3 }}$ & $\n{ 2.5116\cdot 10^{-6 }}$ &
      $\n{ 0.1550}$ & $\n{ 3.1536}$ \\
\hline
Cl---Na &
      $\z{ 9.2873\cdot 10^{-4 }}$ & $\z{ 3.0945\cdot 10^{-6 }}$ &
      $\z{ 0.0166}$ & $\z{ 3.8647}$ \\
Cl---O (S) &
      $\pz{ 6.1201\cdot 10^{-3 }}$ & $\pz{ 1.7946\cdot 10^{-5 }}$ &
      $\pz{ 0.1246}$ & $\pz{ 3.7833}$ \\
Na---O (S) &
      $\pz{ 4.0705\cdot 10^{-4 }}$ & $\pz{ 4.7698\cdot 10^{-7 }}$ &
      $\pz{ 0.0207}$ & $\pz{ 3.2469}$ \\
Cl---O (3) &
      $\z{ 5.9839\cdot 10^{-3 }}$ & $\z{ 1.7342\cdot 10^{-5 }}$ &
      $\z{ 0.1233}$ & $\z{ 3.7759}$ \\
Na---O (3) &
      $\z{ 3.9722\cdot 10^{-4 }}$ & $\z{ 4.5916\cdot 10^{-7 }}$ &
      $\z{ 0.0205}$ & $\z{ 3.2396}$ \\
Cl---O (4) &
      $\z{ 6.0564\cdot 10^{-3 }}$ & $\z{ 1.7592\cdot 10^{-5 }}$ &
      $\z{ 0.1245}$ & $\z{ 3.7773}$ \\
Na---O (4) &
      $\z{ 4.0218\cdot 10^{-4 }}$ & $\z{ 4.6612\cdot 10^{-7 }}$ &
      $\z{ 0.0207}$ & $\z{ 3.2410}$ \\
\hline
\end{tabular}\\[3mm]
\noindent%
\begin{tabular}{lllll}
\hline
\multicolumn{5}{c}{{\normalsize\textbf{OPLS-AA}}} \\
\hline
Atom & $c_6~[\mathrm{kJ}\,\mathrm{nm}^6]$ 
        & $c_{12}~[\mathrm{kJ}\,\mathrm{nm}^{12}]$ & 
        $\epsilon~[\frac{\mathrm{kcal}}{\mathrm{mol}}]$ & $\sigma~[\mathrm{\AA}]$ \\
\hline
Cl &
      $\n{ 1.4654\cdot 10^{-2 }}$ & $\n{ 1.0886\cdot 10^{-4 }}$ &
      $\n{ 0.1178}$ & $\n{ 4.4172}$ \\
Na &
      $\n{ 6.3351\cdot 10^{-5 }}$ & $\n{ 8.6451\cdot 10^{-8 }}$ &
      $\n{ 0.0028}$ & $\n{ 3.3304}$ \\
O (S) &
      $\n{ 2.6188\cdot 10^{-3 }}$ & $\n{ 2.6352\cdot 10^{-6 }}$ &
      $\n{ 0.1554}$ & $\n{ 3.1656}$ \\
O (3) &
      $\n{ 2.4914\cdot 10^{-3 }}$ & $\n{ 2.4367\cdot 10^{-6 }}$ &
      $\n{ 0.1521}$ & $\n{ 3.1506}$ \\
O (4) &
      $\n{ 2.5536\cdot 10^{-3 }}$ & $\n{ 2.5121\cdot 10^{-6 }}$ &
      $\n{ 0.1550}$ & $\n{ 3.1536}$ \\
\hline
Cl---Na &
      $\z{ 1.0227\cdot 10^{-3 }}$ & $\z{ 3.4562\cdot 10^{-6 }}$ &
      $\z{ 0.0181}$ & $\z{ 3.8738}$ \\
Cl---O (S) &
      $\z{ 6.7301\cdot 10^{-3 }}$ & $\z{ 1.9991\cdot 10^{-5 }}$ &
      $\z{ 0.1353}$ & $\z{ 3.7914}$ \\
Na---O (S) &
      $\z{ 4.0810\cdot 10^{-4 }}$ & $\z{ 4.7915\cdot 10^{-7 }}$ &
      $\z{ 0.0208}$ & $\z{ 3.2480}$ \\
Cl---O (3) &
      $\z{ 6.5798\cdot 10^{-3 }}$ & $\z{ 1.9314\cdot 10^{-5 }}$ &
      $\z{ 0.1339}$ & $\z{ 3.7839}$ \\
Na---O (3) &
      $\z{ 3.9820\cdot 10^{-4 }}$ & $\z{ 4.6110\cdot 10^{-7 }}$ &
      $\z{ 0.0205}$ & $\z{ 3.2405}$ \\
Cl---O (4) &
      $\z{ 6.6583\cdot 10^{-3 }}$ & $\z{ 1.9591\cdot 10^{-5 }}$ &
      $\z{ 0.1351}$ & $\z{ 3.7854}$ \\
Na---O (4) &
      $\z{ 4.0311\cdot 10^{-4 }}$ & $\z{ 4.6810\cdot 10^{-7 }}$ &
      $\z{ 0.0207}$ & $\z{ 3.2420}$ \\
\hline
\end{tabular}
\hfill\begin{tabular}{lllll}
\hline
\multicolumn{5}{c}{{\normalsize\textbf{Smith-1994}}} \\
\hline
Atom & $c_6~[\mathrm{kJ}\,\mathrm{nm}^6]$ 
        & $c_{12}~[\mathrm{kJ}\,\mathrm{nm}^{12}]$ & 
        $\epsilon~[\frac{\mathrm{kcal}}{\mathrm{mol}}]$ & $\sigma~[\mathrm{\AA}]$ \\
\hline
Cl &
      $\n{ 1.5790\cdot 10^{-2 }}$ & $\n{ 1.1458\cdot 10^{-4 }}$ &
      $\n{ 0.1299}$ & $\n{ 4.4000}$ \\
Na &
      $\n{ 2.8228\cdot 10^{-4 }}$ & $\n{ 4.7543\cdot 10^{-8 }}$ &
      $\n{ 0.1001}$ & $\n{ 2.3500}$ \\
O (S) &
      $\n{ 2.6172\cdot 10^{-3 }}$ & $\n{ 2.6338\cdot 10^{-6 }}$ &
      $\n{ 0.1553}$ & $\n{ 3.1656}$ \\
O (3) &
      $\p{ 2.4889\cdot 10^{-3 }}$ & $\p{ 2.4352\cdot 10^{-6 }}$ &
      $\p{ 0.1519}$ & $\p{ 3.1508}$ \\
O (4) &
      $\p{ 2.5543\cdot 10^{-3 }}$ & $\p{ 2.5145\cdot 10^{-6 }}$ &
      $\p{ 0.1549}$ & $\p{ 3.1540}$ \\
\hline
Cl---Na &
      $\z{ 2.8223\cdot 10^{-3 }}$ & $\z{ 4.1711\cdot 10^{-6 }}$ &
      $\z{ 0.1140}$ & $\z{ 3.3750}$ \\
Cl---O (S) &
      $\z{ 6.9705\cdot 10^{-3 }}$ & $\z{ 2.0424\cdot 10^{-5 }}$ &
      $\z{ 0.1420}$ & $\z{ 3.7828}$ \\
Na---O (S) &
      $\z{ 9.1847\cdot 10^{-4 }}$ & $\z{ 4.0406\cdot 10^{-7 }}$ &
      $\z{ 0.1247}$ & $\z{ 2.7578}$ \\
Cl---O (3) &
      $\pz{ 6.8132\cdot 10^{-3 }}$ & $\pz{ 1.9730\cdot 10^{-5 }}$ &
      $\pz{ 0.1405}$ & $\pz{ 3.7754}$ \\
Na---O (3) &
      $\pz{ 8.9383\cdot 10^{-4 }}$ & $\pz{ 3.8693\cdot 10^{-7 }}$ &
      $\pz{ 0.1233}$ & $\pz{ 2.7504}$ \\
Cl---O (4) &
      $\pz{ 6.8986\cdot 10^{-3 }}$ & $\pz{ 2.0028\cdot 10^{-5 }}$ &
      $\pz{ 0.1419}$ & $\pz{ 3.7770}$ \\
Na---O (4) &
      $\pz{ 9.0590\cdot 10^{-4 }}$ & $\pz{ 3.9352\cdot 10^{-7 }}$ &
      $\pz{ 0.1245}$ & $\pz{ 2.7520}$ \\
\hline
\end{tabular}\\[3mm]
\normalsize
\caption{Parameters of the Lennard-Jones interactions for different
force fields. The typeface of the numbers indicates where these numbers stem
from. Boldface (``$\g{1.23}$'') means that it is explicitly given by the
force field in the
specified notation. Underlined numbers (``$\n{1.23}$'') denote that 
the Lennard-Jones interaction parameters were given explicitly by the force field 
and a unit conversion (e.\,g. from kcal to kJ) was necessary. 
Normal font (``$\z{1.23}$'') means that
the parameter in question was computed via the combination rule
Eq.~(\protect\ref{eqKombination}). Not all force fields specify all three water
models. In case that one was missing we have taken the missing parameters
(either directly or via the combination rule) from the Gromacs force field. This
is indicated by italic font (``$\pz{1.23}$''). Hydrogens do not participate in
Lennard-Jones interaction, and the symbol after the O for oxygen stands for the
water model (``S'' for SPC and SPC/E, ``3'' for TIP3P, and ``4'' for TIP4P).
}
\label{tabParameter}
\end{table*}

\begin{figure*}
\epsfig{file=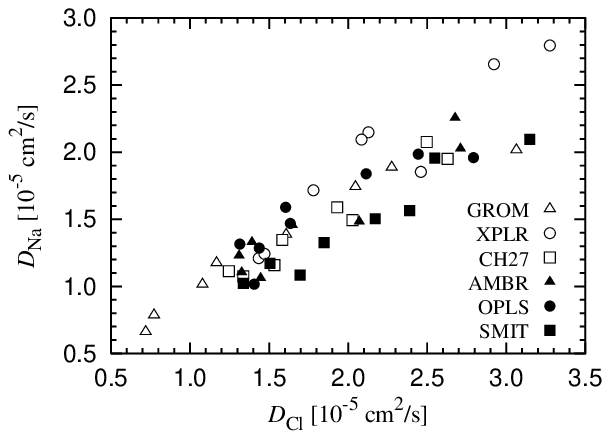,width=3.2in}\hfill\epsfig{file=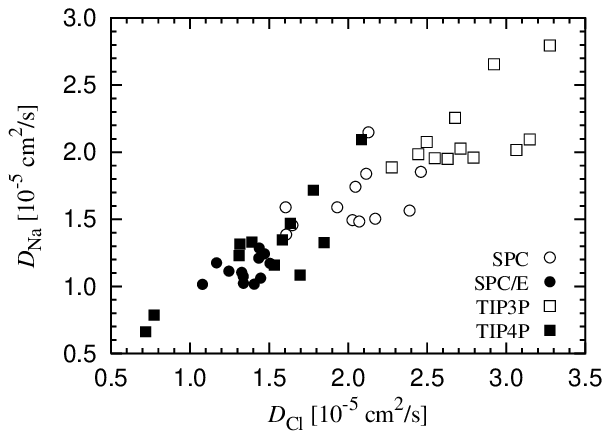,width=3.2in}
\caption{Diffusion coefficients $D_{\mathrm{Na}}$ and $D_{\mathrm{Cl}}$
for sodium and chloride, respectively. 
Left: Labelling according to the ionic force field. Right: Labelling
according to the water model used. 
}
\label{figDiff}
\end{figure*}

In Sec.~\ref{secSimresults} we present the results of our simulations, and show 
how the different force-fields differ in their description of the NaCl
properties. One important conclusion can, however, 
be drawn already from Tab.~\ref{tabParameter}: 
The parameters for the different water models (SPC, TIP3P and TIP4P)
differ only slightly, representing the current good knowledge of the properties of
water. For
(aqueous) chloride, the differences are significantly larger, up to
$10\,\%$ for the radius and up to $50\,\%$ for the depth of the attractive
well of the Lennard-Jones interaction, reflecting the lack of high quality experimental
input data. For (aqueous) sodium, there seems to be virtually
no consensus on its properties. In the simulations, one can thus expect that the
biggest differences will be in the Na--Na properties, followed by the Na--Cl
interactions.

\section{Simulations}

For this study, we decided to include the three most commonly used thermostats,
namely Berendsen~\cite{berendsen:84a}, 
Nos{\'e}-Hoover~\cite{nose:84a,hoover:85a} and Langevin~\cite{grest:86a}. All
of them are implemented into the Gromacs simulation software~\cite{lindahl:01a}
that was used for
all of the computations presented in this paper. The target temperature was set
to $298~\mathrm{K}$ and particle-mesh Ewald (PME) was used for  long-range
electrostatics. The pressure was held constant at $1~\mathrm{bar}$ using the
Berendsen algorithm~\cite{berendsen:84a}.

For each combination of ionic force-field, water model and thermostat a MD
simulation was run. In addition, for each combination of water model and
thermostat a reference simulation without ions was done.
The total number of simulations added up to 84. 
A pre-production analysis 
showed that the systems needed slightly less than $0.5~\mathrm{ns}$ to
equilibrate. For each simulation run, we computed a $2~\mathrm{ns}$
trajectory and only the second half of that was included into the analysis.

The simulations were run at the physiological salt concentration of
$0.87~\mathrm{mol}$. The simulation box contained slightly more than 10000
water molecules so that finite size effects are not expected.
Lennard-Jones interaction was cut-off at $1~\mathrm{nm}$. The optimal choice
for the cutoff length is not obvious and can vary between force fields (even
between the ones for the ions and for water in the same simulation). 
For consistency, we decided to use the
same cutoff in all simulations. For all of these systems,
all relevant structures are on scales much smaller than
$1~\mathrm{nm}$. Furthermore, all atoms are charged so
that Lennard-Jones interactions quickly become negligible compared to
electrostatic interactions, and the precise choice of cutoff does not matter as
much as it does for other systems.

The simulations described above presented a significant numerical task, and a total
of approximately 25\,000 hours of cpu time was needed to complete them.

\section{Simulation results}
\label{secSimresults}

\subsection{Dynamic properties}

The most common quantity to describe the dynamical behaviour of a system is its
diffusion coefficient $D$. We have plotted the results for different 
forcefields and water models in Fig.~\ref{figDiff}. The format for
this plot, as well as the following ones, is the following:
All results are plotted twice, using two graphs next to each other. 
The data points in both graphs are identical but in the left figure we
have labelled them (i.\,e., picked symbols for the data points) 
according to the ionic force field whereas
in the right figure we have labelled them according to the
water model used. This way it is easy to see whether 
there is any systematic dependence on the ionic force field and/or the water model.

\begin{figure*}
\epsfig{file=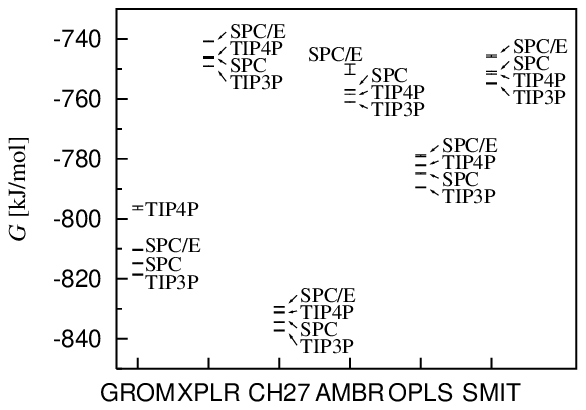,width=3.2in}\hfill\epsfig{file=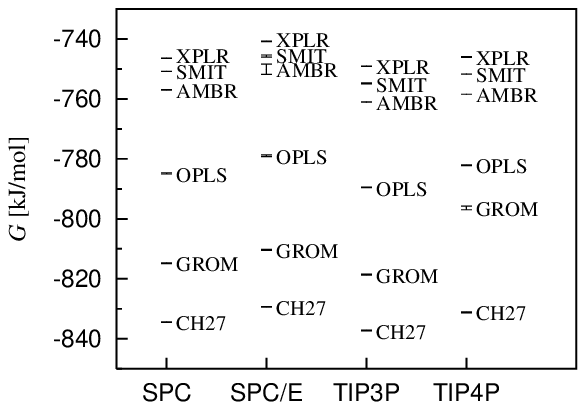,width=3.2in}
\caption{Gibbs free energy of hydration for the
different force fields per ion pair. }
\label{energieFig}
\end{figure*}

The results for diffusion coefficients using 
Berendsen  and Nos{\'e}-Hoover were identical within statistical error,
but using  Langevin
thermostat the diffusion coefficients were much smaller. 
Unlike the Berendsen  and Nos{\'e}-Hoover thermostats, the Langevin thermostat
is not momentum conserving, and thus 
we omit the results for the Langevin
thermostat when computing the diffusion coefficients.

The results of our simulations are depicted in Fig.~\ref{figDiff}.
The experimental values at infinite dilution are given by
$D_{\mathrm{Na}}=1.334\cdot 10^{-5}~\mathrm{cm}^2/\mathrm{s}$
and $D_{\mathrm{Cl}}=2.032\cdot 10^{-5} ~
\mathrm{cm}^2/\mathrm{s}$~\cite{marcus:97}. 
From the figure it is seen immediately that
the dynamics is
determined by the water model while the contribution of the ionic force field
is negligible. The diffusion constant thus cannot be used to judge the
quality of the different force fields. This
is not surprising since in aqueous systems the behaviour is dominated by the
water molecules as they outnumber the ions by a factor of order $100$. This is
likely to be the case for other dynamic properties as well. A
study of the dynamic properties of water models is beyond the scope and aim of this
paper, and we refer to previous studies on this subject~\cite{wallqvist:99a}
and to the very informative
webpage~\footnote{http://www.sbu.ac.uk/water/models.html}. To see the effect of
the ionic force fields, we concentrate on static properties in the following.

\subsection{Gibbs free energy of hydration}
\label{secEnergy}

The most frequently used way to compute free energies is by the ``slow-growth''
method, also known as Kirkwood's coupling parameter method~\cite{kirkwood:35a} 
or as thermodynamic integration.
The presence of solute in a solvent is determined by the parameter
$\lambda$ that modifies the Hamiltonian $H$ of the system. $\lambda=0$ means that
there is no interaction between solute and solvent whereas $\lambda=1$ means
that the normal Hamiltonian for the combined system is used. The change of
Gibbs free energy upon introducing the solute into the solvent is then given by
\begin{equation}
        \Delta G = \int_0^1 \left\langle \frac{\partial H}{\partial \lambda}
                \right\rangle d\lambda \;,
\end{equation}
where the averages are taken at constant pressure. (If the averages are taken at
constant volume, this equation would yield the change of free energy.)
The advantage of this
approach is that the change of (Gibbs) free energy is measured directly
without the need to
compute the (Gibbs) free energy of the solvent, thereby reducing the statistical
error of the result.
However, this method is computationally expensive since the
integral has to be evaluated numerically by running simulations with different
values of $\lambda$. While this is no significant problem when
studying a single system, this is not feasible for a systematic study as
presented in this paper.

A different way to compute the free energy of a system is given by
\begin{equation}
        F = - k_{\mathrm{B}} T \ln \Bigl\langle 
                \exp\bigl( - \frac{U}{k_{\mathrm{B}} T} \bigr) \Bigr\rangle \;,
        \label{eqFree2}
\end{equation}
with $U$ the potential energy of the system. This formula is easily understood
by noting that it is conceptually identical to the computation of the chemical
potential using insertion of a test particle, with the test particle being the
entire system and the reference state being the vacuum. It offers the 
advantage that only
a single simulation of the system with and without the solute is needed,
not a large number of different simulations with gradually increasing $\lambda$.
This formula is not as much used
in the literature since for computing free energy differences it has to be
applied twice to compute the free energies of the two comparison systems. The
computed free energy difference then has a significantly larger statistical
error, especially
when only a small part of, for example, a large protein is mutated.

\begin{figure*}
\epsfig{file=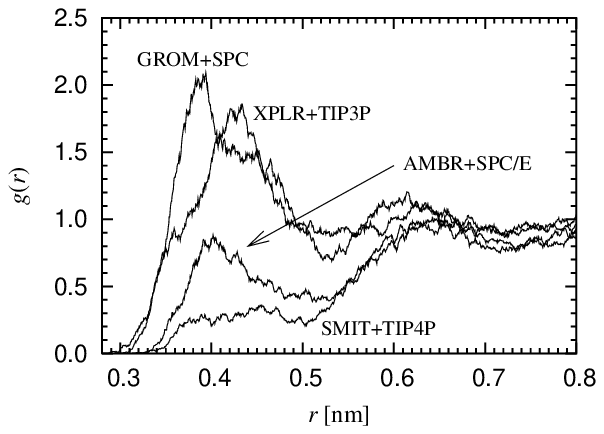,width=2.3in}\begin{picture}(0,0)
\put(-45,96){Na--Na}\end{picture} \hfill
\epsfig{file=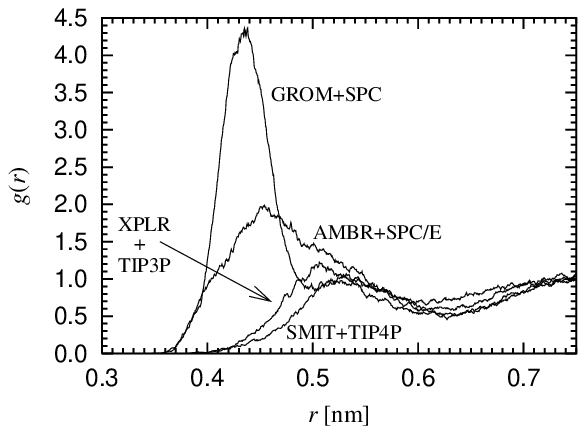,width=2.3in}\begin{picture}(0,0)
\put(-45,96){Cl--Cl}\end{picture} \hfill
\epsfig{file=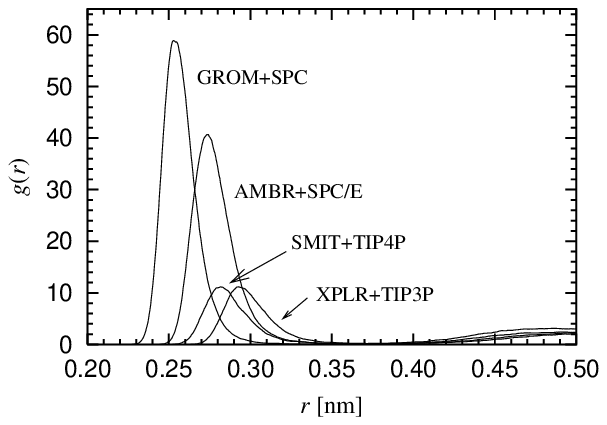,width=2.3in}\begin{picture}(0,0)
\put(-45,96){Na--Cl}\end{picture}
\caption{Typical examples from our set of radial-distribution functions for
Na--Na, Cl--Cl and Na--Cl (from left to right). All curves were computed using
the Berendsen thermostat.}
\label{figRdfBeispiel}
\end{figure*}

We do not suffer from this problem as the number of ions is large enough to
arrive at a statistically significant result. We thus apply Eq.~(\ref{eqFree2})
first to a simulation of aqueous NaCl and then to a simulation with the same
number of water molecules (using the same water model and the same thermostat
as in the first simulation) but without any ions. Since the two simulations
are done at identical pressure and not at identical volume, the difference
actually gives not the plain free energy but rather the Gibbs free 
energy $G$ of hydration.

The experimental values for the Gibbs free energy of hydration are
$-347~\mathrm{kJ}/\mathrm{mol}$ for chloride and
$-375~\mathrm{kJ}/\mathrm{mol}$ for sodium~\cite{marcus:97}, for a total of
$-722~\mathrm{kJ}/\mathrm{mol}$. Older reported values are
$-363~\mathrm{kJ}/\mathrm{mol}$ for chloride and
$-406~\mathrm{kJ}/\mathrm{mol}$ for sodium~\cite{cotton:72}.
It is assumed that isolated
ions are to be hydrated, i.\,e., the crystal lattice of solid NaCl does no
longer need
to be broken up. (These two different concepts are frequently referred to as
``hydration'' versus ``solvation''.) 
It should be stressed that all those experimental
values are at infinite dilution, i.\,e., all mutual interactions among the
ions in the aqueous phase are ignored. 

The numerically computed Gibbs free energies $G$ are depicted in
Fig.~\ref{energieFig}. Since we used three different thermostats, this resulted
in three different values for $G$ for each combination of ionic forcefield
and water model. 
We use those three different values to compute an error estimate
that is marked by the error bar in the figure. Most of the error bars are that
small that they hardly can be seen, confirming that application 
of Eq.~(\ref{eqFree2}) is
sufficient for this system.

The experimental data given above for the Gibbs free energy of hydration is at
infinite dilution and thus needs to be corrected to account for the finite
molarity of physiological systems as treated in the simulations. We were unable
to find direct experimental data for the enthalpy change of dilution but this
quantity can be computed from the difference in the enthalpy of solution for the
two different concentrations. For the parameters in question, the experimental
values~\cite{beggerow:76} lead to a
correction of $-2~\mathrm{kJ}/\mathrm{mol}$.

All numerically computed data in Fig.~\ref{energieFig} are more negative than
the experimental data. This means that either the attractive mutual interaction
of the ions is overestimated in the simulations, or that the shielding of that
interaction due to the water is underestimated. It should be stressed that the
parametrisation of all ionic force fields was done at infinite dilution,
meaning that only a single ion was considered. The effect of finite
concentration was thus not considered in the parametrisation process. A
systematic numerical study of the concentration dependence of the Gibbs free
energy of hydration could help to provide an understanding but no such study
seems to have been published so far. Comparing experiment and the simulation
results in Fig.~\ref{energieFig} allows to judge the agreement between a force
field and experiment --- for precisely the molarity studied in this
paper --- but the picture might be completely different at other
molarities.

\subsection{Radial-distribution functions}

Next, we compare the  radial-distribution functions (rdf). There are three
kinds of particles in the simulation, namely Na, Cl and water, resulting in six
different pairs and thus six different radial-distribution functions. For the
purpose of computing the rdf's, the position of the oxygen atom is taken to
represent the entire water molecule, and we will thus use the label ``O'' for
water. For space reasons we will not discuss the
water-water rdf's in the following. This is no relevant limitation since for
the physiological concentrations they depend hardly on the ionic force  field.
(The complete set of rdf's can be found as supplementary
material at our website~\footnote{http://www.softsimu.org/biophysics/ions/}.)

In Fig.~\ref{figRdfBeispiel} we
present the rdf's for four different force field combinations. (For the
complete set, see the supplementary material.) It is
immediately obvious that the rdf's differ from each other 
in many aspects, such as  the number of peaks,
the relative and absolute heights of the peaks, 
and that those differences are significant.

\begin{figure*}
\centering
\begin{tabular}{@{}ccc@{}}
\epsfig{file=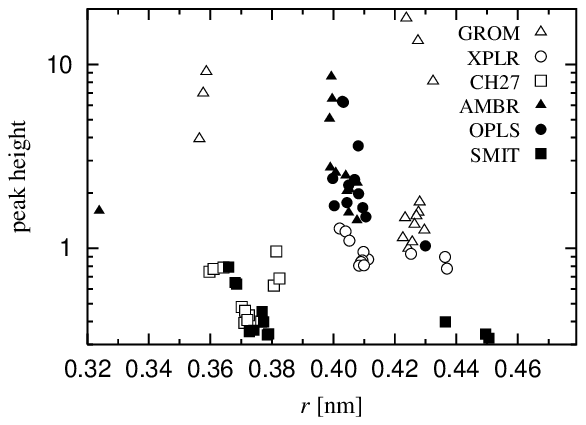,width=2.8in} 
        & \epsfig{file=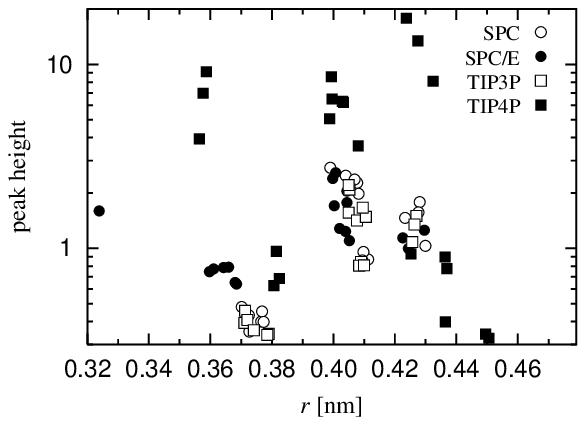,width=2.8in} 
        & \raisebox{2.2cm}{\rotatebox{90}{\textbf{Na--Na}}}\\
\epsfig{file=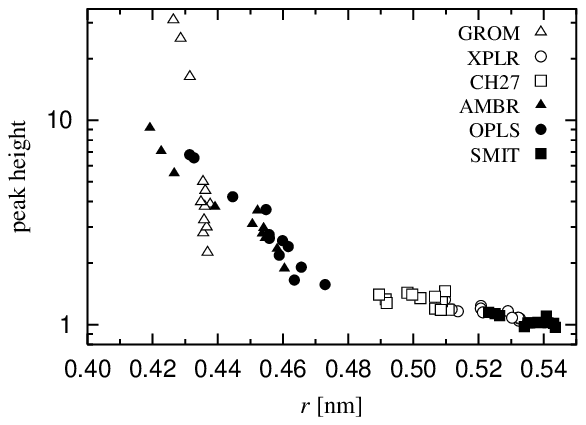,width=2.8in} 
        & \epsfig{file=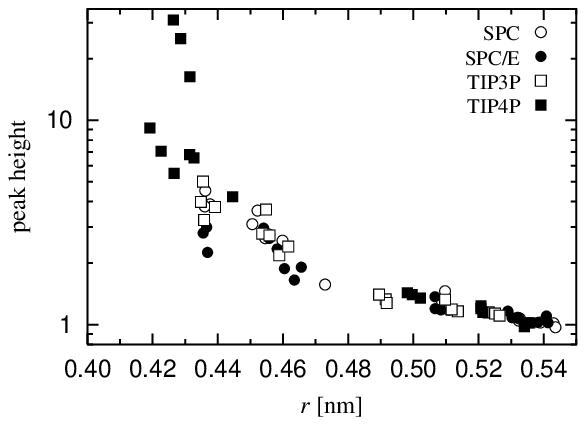,width=2.8in} 
        & \raisebox{2.2cm}{\rotatebox{90}{\textbf{Cl--Cl}}} \\
\epsfig{file=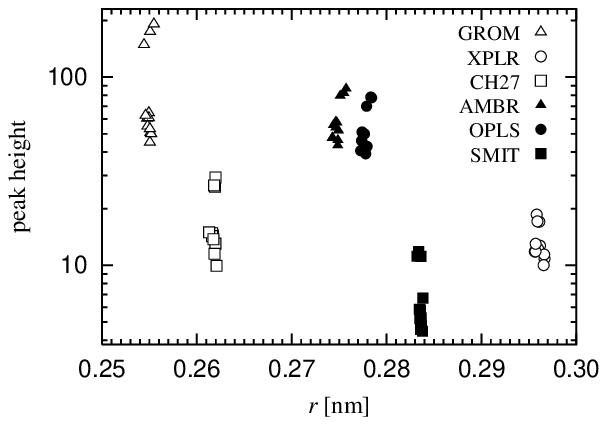,width=2.8in} 
        & \epsfig{file=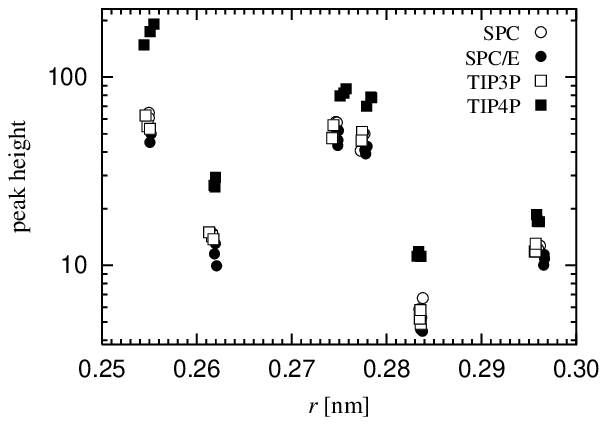,width=2.8in} 
        & \raisebox{2.2cm}{\rotatebox{90}{\textbf{Na--Cl}}} \\
\epsfig{file=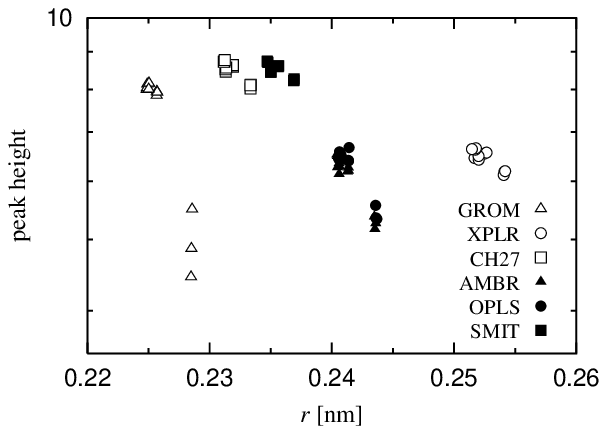,width=2.8in} 
        & \epsfig{file=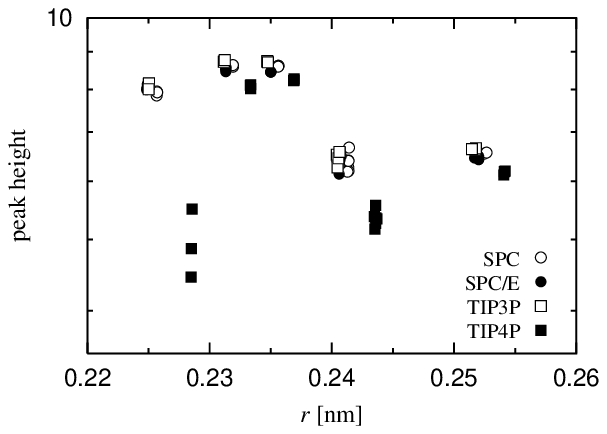,width=2.8in} 
        & \raisebox{2.3cm}{\rotatebox{90}{\textbf{Na--O}}} \\
\epsfig{file=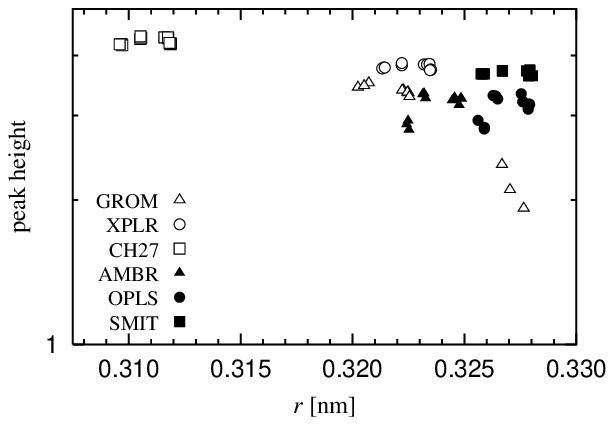,width=2.8in} 
        & \epsfig{file=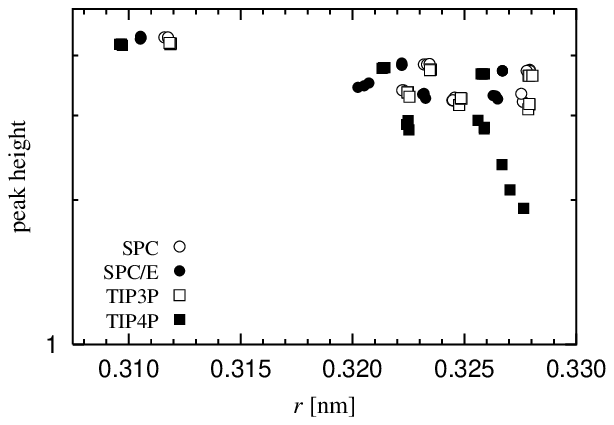,width=2.8in} 
        & \raisebox{2.3cm}{\rotatebox{90}{\textbf{Cl--O}}}
\end{tabular}
\caption{Position and height of the first peak of the radial-distribution
function for (from top to bottom) Na--Na, Cl--Cl, Na--Cl, Na--O (i.\,e., water)
and Cl--O.}
\label{rdfFig}
\end{figure*}

It would be impossible to present all rdf's by directly plotting them.
To give a more systematic overview in a condensed way, we have computed the
position and height of the first peak for all rdf's. 
For this, a Gaussian is fitted to the rdf in the neighbourhood of the peak.
The results are depicted in Fig.~\ref{rdfFig}. We first will discuss the
simulation results and will then put them into the context of experimental
results.

\begin{figure*}
\epsfig{file=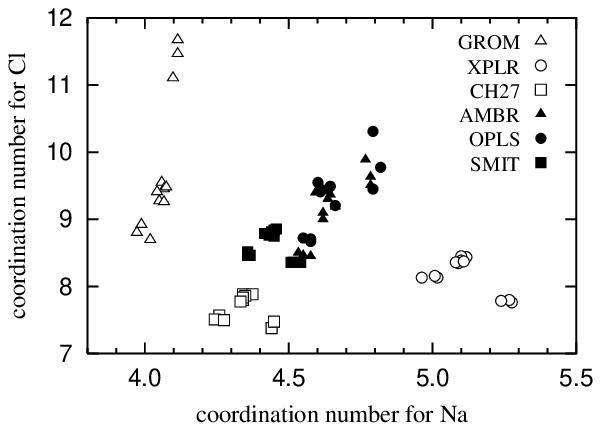,width=3.2in}\hfill%
\epsfig{file=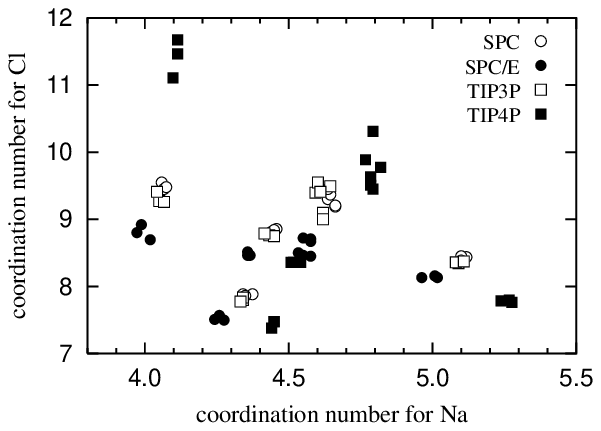,width=3.2in}
\caption{Coordination number, computed by integrating the
radial-distribution function between the water oxygen and the ion 
from zero up to its first minimum.}
\label{figCoord}
\end{figure*}

From Fig.~\ref{rdfFig} it is seen that the Na--Na and Cl--Cl peaks
are scattered widely (Na--Na being scattered more than Cl--Cl), and that there
is no well-defined systematic tendency. This is the case for both the position
and the height of the peak. Also for the Na--Cl results, the peaks are scattered
widely but we want to point out one curiosity: While for the Na--Na and Cl--Cl peaks there is a dependence both on the
ionic force field and the water model, the position of
the Na--Cl peaks is independent of the chosen water model. An easy explanation
could be that sodium and chloride ions are frequently in direct contact with 
each other so that the properties of the water model are of minor
importance. 

The positions of the Na--O and Cl--O peaks differ less between force
fields. (Please observe the
different scale of the $r$-axis in the different subfigures.) The height of the
peaks, in contrast, still varies by about a factor $3$ between the different
force fields.

We will now discuss these findings in the context of experimental limitations.
Radial-distribution functions can, in principle, be measured by x-ray
diffraction since they are related to the Fourier transformation of the
structure factor. To get a strong enough signal, the substances have to be of
sufficiently high concentration. This makes the experimental determination of
Na--Na, Na--Cl or Cl--Cl rdf's impossible.

For Na--O and Cl--O rdf's, the signal is only weak at physiological
concentrations.  It is possible to determine the position of
the peak of the rdf quite well since it is directly related to the wave length
of the peak of the structure factor. The height of the peak of the rdf,
however, can be measured only with great difficulty. (We will return to this in
the next section, when discussing the coordination number.) We should add that
for sodium any kind of measurement is more difficult than for chloride since
there exists only a single isotope of sodium, and consequently the difference
method of isotope substitution cannot be used.

Determining suitable experimental values for the peak positions is aided by the
experimental observation that they depend only very weakly on molarity of the
salt. Surprisingly, the results for crystals and aqueous solutions are also
very similar. Marcus~\cite{marcus:97} quotes values between $0.233~\mathrm{nm}$
and $0.240~\mathrm{nm}$ for the Na--O peak position but earlier reviews include
values up to $0.25~\mathrm{nm}$~\cite{marcus:88a}. For chloride, the quoted
values are between $0.30~\mathrm{nm}$ and $0.32~\mathrm{nm}$~\cite{marcus:97}.
Other values quoted in the literature are within these limits, except that also
a Cl--O peak position of up to $0.335~\mathrm{nm}$ is
reported~\cite{enderby:79a}.

The simulation results in Fig.~\ref{rdfFig} are now easily understood. There 
are only two quantities that are susceptible to (decent-quality) measurement,
namely the peak positions of the Na--O and Cl--O rdf's. Consequently the
predicted values for these two quantities vary only little between force
fields. Taking the entire range of experimental values (i.\,e., not deciding by
some criterion that one experimental value is ``better'' than the others), all
force fields basically reproduce the experimental knowledge.

The huge spread observed in the other peak positions and in all the
peak heights
in Fig.~\ref{rdfFig} is simply an echo of the lack of
experimental data for these important structural quantities, and no empirical
force field can be better than the available experimental data. Any method that
relies on mutual peak position of the ions therefore faces a problem, and no
solution to this problem is in sight.

\subsection{Coordination number}


A different description of the water structure around the ions is given by the
coordination number. Technically this quantity is computed by first determining
the location of the first minimum of the radial-distribution function and then
integration the radial-distribution function from zero up to this point,
\begin{equation}
n_{\mathrm{coord}} = \frac{4 \pi}{ V_{\mathrm{water}}} \int_0^{r_{\mathrm{min}}}
        d r \, r^2 g(r) \;,
        \label{eqCoord}
\end{equation}
where $V_{\mathrm{water}}$ is the volume of one water molecule.
The
geometric meaning of $n_{\mathrm{coord}}$ is that it gives the number of
solvent molecules in the first hydration shell around the ion. It should be
noticed that the coordination number is a purely geometrical quantity, and it
does not imply that this number of water molecules is influenced in some way by
the ion.

\begin{figure*}
\epsfig{file=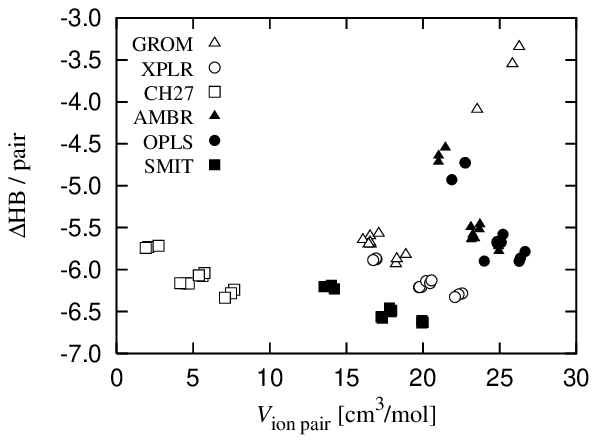,width=3.2in}\hfill\epsfig{file=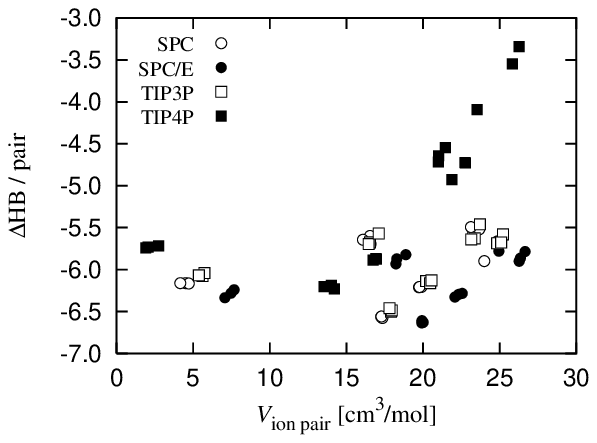,width=3.2in}
\caption{Apparent molar volumes of the ions, and the change of
the number of hydrogen bonds per ion pair. (The latter quantity is negative,
meaning that hydrogen bonds are destroyed by the ions.)}
\label{figHbond}
\end{figure*}

Experimental data on the coordination number is scarce. The reason for this is twofold:
First, the amplitude of the radial-distribution function is difficult to
measure. The amplitude of $g(r)$ cannot directly be inferred from the structure
factor at the corresponding wave length but only follows indirectly from the
normalisation condition $g(r)=1$ for $r\to\infty$. Secondly, the position
of the first minimum is less well-defined than the position of the first peak,
and it is not always immediately obvious where to stop the integration. (This
can already be seen from the examples of simulational rdf's in
Fig.~\ref{figRdfBeispiel}).

The coordination numbers computed from our simulations are displayed in
Fig.~\ref{figCoord}. First, the position of the first minimum was computed.
For space reasons we decided against displaying these intermediate results and
only show the resulting coordination numbers.
(The positions of the first minimum
are between $0.305~\mathrm{nm}$ and $0.34~\mathrm{nm}$ for Na--O, and
between $0.375~\mathrm{nm}$ and $0.44~\mathrm{nm}$ for Cl--O.)

The reported experimental coordination numbers for sodium are between $4$ and
$8$~\cite{marcus:88a}. For chloride,  the same source gives values for chloride
of around $6$ when measured for NaCl solutions but there are only few of those
measurements. The coordination number of chloride should change only slightly
when the sodium is replaced by some other cation. This should allow us to use
also data for other salts, thereby getting better statistics. However, if this
is done coordination numbers as high as $11$ can be included~\cite{enderby:79a}.
Given the large
spread of the experimental values, the coordination number is not well suited
to quantify the agreement between experiment and molecular dynamics
simulations.

\begin{figure*}
\epsfig{file=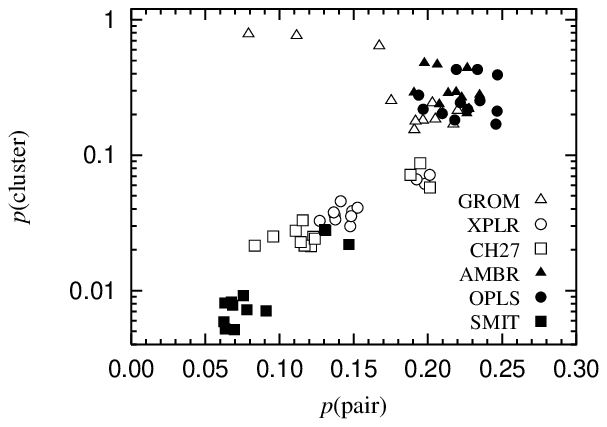,width=3.2in}\hfill\epsfig{file=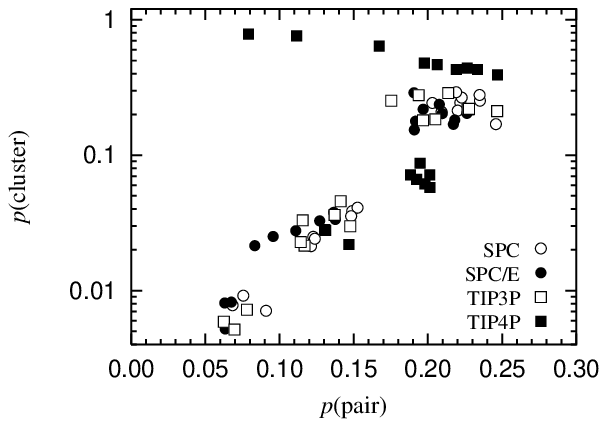,width=3.2in}
\caption{Ratio $p(\mathrm{pair})$ of all particles that are in a cluster
consisting of two particles, and ratio $p(\mathrm{cluster})$ of particles that
are in a cluster consisting of three or more particles.}
\label{figCluster}
\end{figure*}

\begin{figure*}
\epsfig{file=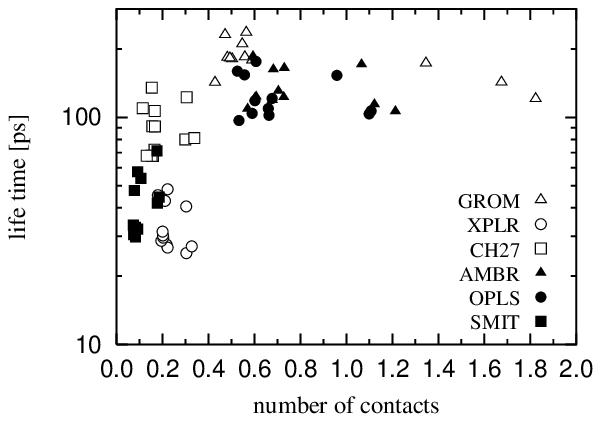,width=3.2in}\hfill\epsfig{file=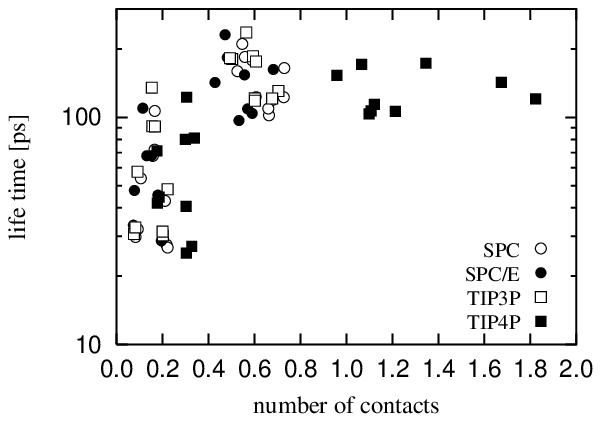,width=3.2in}
\caption{Life time of a contact between a cation and an anion, as well as the
mean number of contacts that each ion has.}
\label{figContact}
\end{figure*}

\subsection{Molar volumes}

Each ion occupies a certain volume determined by its radius. This quantity is
of little practical relevance, however, since the introduction of the ion
changes the local structure of the surrounding water, and thus also the local
density. These effects are included in the quantity known as molar volume. If
the local compression of the water overcompensates the steric volume of the
ion, the molar volume can become negative. 

The experimental molar volumes are $-6.7~\mathrm{cm}^3/\mathrm{mol}$ for sodium
and $23.3~\mathrm{cm}^3/\mathrm{mol}$ for
chloride~\cite{millero:71a,akitt:80a,marcus:97}.
This gives a volume
of $16.6~\mathrm{cm}^3/\mathrm{mol}$ per ion pair.

In a simulation using pressure coupling, the molar volume per ion pair can
directly be computed by determining the average size of the simulation box, and
from this subtracting the average size of the simulation box of a simulation of
pure water. (It is essential to use the same water model and the same thermostat
in both simulations.) This gives the molar volume per ion pair depicted in
Fig.~\ref{figHbond}. Comparison with the experimental values shows
that Charmm-27, OPLS and Amber are
performing a little bit less well in this test as do the other force fields.

\subsection{Hydrogen bonding}

Most ions are known to break the hydrogen bond network of the water
around them. This can be
explained by the water molecules aligning themselves for better interaction with
the ion, at the expense of breaking the hydrogen bonds toward other water
molecules. (On average, a water molecule has about $1.55$ hydrogen bonds in bulk
water.) Experimental values for $\Delta\mathrm{HB}$, the change
in the number of hydrogen bonds due to one ion, is
$-0.03$ for sodium and $-0.61$ for chloride~\cite{marcus:94a,marcus:97}.

We have computed the change in the number of hydrogen bonds by comparing two
simulations, one with and one without the ions. Classical MD can only give an
estimate of the number of hydrogen bonds, by counting the number of potential
acceptors and donors that fulfil certain geometric conditions. The result is
depicted in Fig.~\ref{figHbond}. It is immediately seen that all forcefields
yield values for $\Delta\mathrm{HB}$ that are much more negative than the
experimental values.

An easy explanation can be offered for this observation. All force fields
discussed in this paper do not include explicit energy terms for hydrogen
bonds. Rather, hydrogen bond interaction is integrated out and is included
in the partial charges inside the water molecule. This approach
works very well if the water molecules are arranged in the same way as in bulk
water. In any other arrangement, e.\,g. if ions are introduced, they are now
easier to rotate as no hydrogen bond energy terms need to be broken.

\subsection{Cluster analysis}

Next we discuss the physical background behind the peaks in the  ionic rdf's in
Fig.~\ref{rdfFig}. For the Na--Cl interaction a large peak is expected since
the two ions are of opposite charge. For the Na--Na and Cl--Cl rdf's, this
explanation does not hold. Especially if the peak is significantly higher than
$1$, this means that ions ``like'' to be at a certain distance from each other
much more than to be away from each other as much as possible --- even though
they are strongly repelling each other through electrostatic forces. In earlier
simulations of aqueous ionic systems pairing of chloride ions was found but it
was later realised that this pairing is an artifact that disappears if
long-range electrostatic forces are treated
properly~\cite{perera:94a,buono:94a,hummer:93a}. 

Such problems can be ruled out here but there still is the question whether
clusters of ions exist. In such a cluster, several cations (anions) can be close
to each other because their mutual repulsion is compensated by the 
simultaneous presence of
several anions (cations).
For this reason we have performed a cluster analysis. 
We define a cluster as the set of all ions that are connected by distances of
$0.35~\mathrm{nm}$ or less. From the radial-distribution functions in
Fig.~\ref{rdfFig} it can be seen that $0.35~\mathrm{nm}$ is a reasonable value,
and that the precise choice does not effect the results.

\begin{figure*}
\epsfig{file=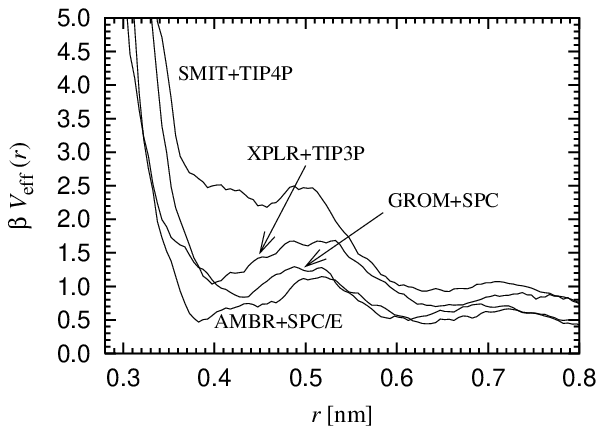,width=2.3in}\begin{picture}(0,0)
\put(-45,96){Na--Na}\end{picture} \hfill
\epsfig{file=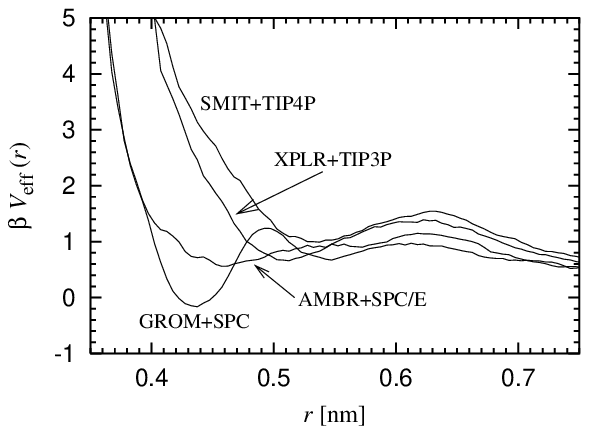,width=2.3in}\begin{picture}(0,0)
\put(-45,96){Cl--Cl}\end{picture} \hfill
\epsfig{file=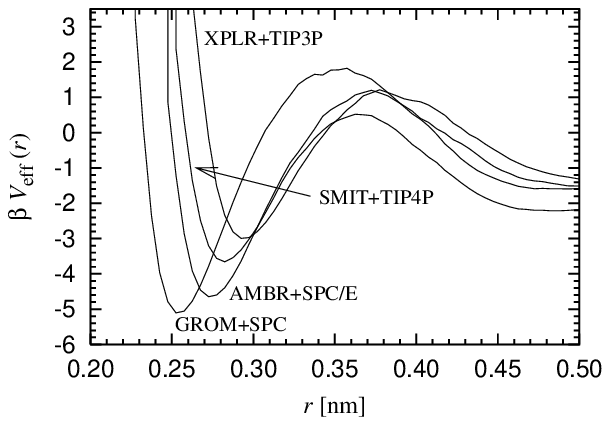,width=2.3in}\begin{picture}(0,0)
\put(-45,96){Na--Cl}\end{picture}
\caption{Effective potentials for the same simulations as depicted in 
Fig.~\protect\ref{figRdfBeispiel}. The product of the effective potential
$V_{\mathrm{eff}}$ and the Boltzmann factor $\beta\equiv 1/(k_{\mathrm{B}} T)$
is unitless.}
\label{figEffBeispiel}
\end{figure*}

After doing this, we collected statistics on the number of ions in each cluster,
computing the ratio $p(N)$ of ions that belong to a cluster consisting of
$N$ ions.
In Fig.~\ref{figCluster} we have plotted
$p(\mathrm{pair})\equiv p(2)$ and $p(\mathrm{cluster})\equiv \sum_{n=3}^{\infty}
p(n)$. It can be seen that the different forcefields result in a large span for
the results, and the probability for clusters of three or more particles spans
more than two orders of magnitude.

The different tendency for cluster formation is reflected also in the mean
number of contacts that each ion has. This quantity is depicted in
Fig.~\ref{figContact}. Also depicted in this figure is the mean life time of
each contact. This shows that if the life time is small, the mean number of
contacts is also small but a large life time does not necessarily imply a large
number of contacts. This can be explained by the interplay between the time
scale for breaking a contact (i.\,e., the life time) and the time scale needed
by two partners to find each other.

\subsection{Implications for the effective potential}

The radial-distribution function can be used to define different potentials. The
most common one is the potential of mean force $V_{\mathrm{PMF}}$, defined
by
\begin{equation}
        g(r) = \exp[-\beta V_{\mathrm{PMF}}(r)] \;.
        \label{eqPMF}
\end{equation}
Although not immediately visible in the formula, 
the potential of mean force  includes the direct interaction between 
two particles at fixed positions, and additionally the contribution 
from having a third particle at a fixed position provided particles 
1 and 2 are already fixed~\cite{hansen:86}. 
In other words, the potential of mean force includes first order corrections
to the pure pairwise potential.
Differences in $g(r)$ directly translate
into differences in $V_{\mathrm{PMF}}(r)$.


If higher order corrections are included, a different kind of potential
is found, termed effective potential
$V_{\mathrm{eff}}(r)$~\cite{lyubartsev:95a}. The effective potential
is defined by the condition that in the canonical ensemble it yields the
desired radial-distribution function. Under the restriction that the Hamiltonian
can be written as a sum of two-particle terms, the effective potential always
exists and is uniquely defined (up to a physically irrelevant offset). If the
full set of rdf's is used for this process, the original microscopic potential
is recovered.

The value of the effective potential lies in that it allows one to integrate out
degrees of freedoms. This process is referred to as coarse-graining. Of
particular interest is to integrate out the solvent since this decreases the
number of particles and thus the computational burden by approximately a factor
$100$. Aim is thus to find three effective potentials, namely for Na--Na,
Na--Cl and Cl--Cl interaction, that,  when used in a MC or MD simulation
without a solvent, reproduce the microscopic rdf's for Na--Na, Na--Cl and
Cl--Cl computed including the solvent. 

As initial guess for $V_{\mathrm{eff}}(r)$, the potential of mean force from
Eq.~(\ref{eqPMF}) is used. From the guessed effective potentials, the rdf's
are  computed in an MC simulation. If the rdf for particles of type $i$ and $j$
at distance $r$ is larger (smaller) than the microscopic target rdf, the guess
for $V^{(i,j)}_{\mathrm{eff}}(r)$ is decreased (increased). This process is
repeated until convergence is achieved. More sophisticated methods that use the
four-particle correlation function to compute an improved guess for
the effective potential have been developed under the name inverse
Monte-Carlo~\cite{lyubartsev:95a} but they are not numerically stable enough
for this problem.

The effective potential for the rdf's from Fig.~\ref{figRdfBeispiel} are shown
in Fig.~\ref{figEffBeispiel}. It is instructive to compare these two figures
since it demonstrates the difference between the potential of mean force and
the effective potential. The Cl--Cl radial-distribution functions of both
Gromacs+SPC and X-plor+TIP3P have a peak.  According to Eq.~(\ref{eqPMF}) this
translates into a dip in  $V_{\mathrm{PMF}}$. For $V_{\mathrm{eff}}$, this dip
is almost invisible for X-plor+TIP3P. This means that the peak in $g(r)$ is
(almost) solely due to interaction with particles that were not integrated out.
In other words: Two chloride ions have a tendency to be close to each other due
to attractive  interaction with the same sodium ion. For Gromacs+SPC,
$V_{\mathrm{PMF}}$ contains a big dip for the Cl--Cl interaction. This means
that a big contribution to the peak in $g(r)$ comes from interactions that have
been integrated out, namely the water molecules.

The important conclusion, however, is that not only the structure of the ionic
solution, as given by $g(r)$, depends heavily on the force field used. Also the
effective potentials show a large variation on the precise force field
parameters.

\section{Conclusions}

In this paper, we have compared different force fields describing aqueous
NaCl.  To this end, we have computed several quantities, some of them of a
thermodynamic origin but being of structural nature. We can divide the
computed quantities into two groups. In the first group are the quantities that
have an experimental counterpart that is easily measurable. All force fields
reproduce most of those values sufficiently well. In the second group are the
quantities for which there exists no or very limited 
experimental data. The different force fields predict vastly 
different values for those quantities.

The Gibbs free energy of hydration is the most basic  thermodynamic quantity
for describing aqueous ionic solutions. It is a ``purely thermodynamic''
quantity if and only if the system is in the dilute limit. For finite
concentration of ions, the mutual interaction of ions becomes important, and
thus the structural properties of the ions enter. Comparison of experimental
and our simulational values for the Gibbs free energy of hydration shows that
the difference between those values depends on the force field used, so in
principle it can be used to judge the suitability of a force field for
application at physiological salt concentrations. While the Gibbs free energy
of hydration is an important fundamental thermodynamic property, its
importance in most molecular dynamics simulations might be limited,
however, since the
ions are usually kept hydrated during the entire simulation.

Structural properties are often more important, best described by the
radial-distribution function.
The position of the first peak of the radial-distribution functions for
sodium--water and chloride--water is characteristic for the first hydration
shell of those ions. All force fields reproduce the experimental values
sufficiently well. This means that all force fields are able to describe well
those physical effects that are governed by the hydration properties in the
immediate neighbourhood of the ions.

For the ion--ion radial-distribution functions it was shown that the different
force fields lead to significantly different results. Since there exists no
experimental data for those properties, this does not imply that a few (or all)
force fields were developed improperly but simply reflects the limited
knowledge of such properties. The predictive power of any model
for aqueous NaCl thus is
very limited if structural properties of mutual ion interaction are important.
In particular this means that greatest care should be taken
if the radial-distribution functions are used as 
input for some other application.

\acknowledgments

This work has been supported by the European Union Marie Curie fellowship
HPMF-CT-2002-01794 (M.\,P.) and
by the Academy of
Finland grant no.~54113 (M.\,K.).

\end{document}